\documentclass[aps,prd,preprintnumbers,showpacs,nofootinbib,twocolumn]{revtex4}
\usepackage{graphicx}
\usepackage{dcolumn}
\usepackage{bm}
\usepackage{epsfig} 
\usepackage{subfigure}
\usepackage[english]{babel}
\usepackage{amsmath}
\usepackage{amssymb}
\usepackage{enumerate}
\usepackage{graphics,psfrag,rotating}
\usepackage{float} 
\usepackage[utf8]{inputenc}
\usepackage[english]{babel}
\usepackage{amsmath}
\usepackage{amsfonts}
\usepackage{amssymb}
\usepackage{epstopdf}
\usepackage{color}
\usepackage{hyperref}
\usepackage{appendix}

\def\be{\begin{equation}}
\def\ee{\end{equation}}
\def\bea{\begin{eqnarray}}
\def\eea{\end{eqnarray}}

\begin{document}

\title{Two new tests to the distance duality relation with galaxy clusters}

\author{Simony Santos-da-Costa$^{a,b}$
Vinicius C. Busti$^{c,d}\,\footnote{vcbusti [at] astro.iag.usp.br}$
Rodrigo F. L. Holanda$^{e,b}$
}
\affiliation{
$^{a}$ Departamento de Astronomia, Observat\'{o}rio Nacional,
    Rio de Janeiro, RJ, Brazil 20921-400\\
$^{b}$ Departamento de F\'{\i}sica, Universidade Federal de Campina Grande, 58429-900, Campina Grande - PB, Brazil \\
$^{c}$  Astrophysics, Cosmology and Gravity Centre (ACGC), Department of Mathematics and Applied Mathematics, University of Cape Town, Rondebosch 7701, Cape Town, South Africa\\
$^{d}$ Departamento de F\'{i}sica Matem\'{a}tica, Instituto de F\'{i}sica, Universidade de S\~{a}o Paulo, CP 66318, \\
CEP 05508-090, S\~{a}o Paulo - SP, Brazil \\
$^{e}$ Departamento de F\'{\i}sica, Universidade Estadual da Para\'{\i}ba, 58429-500, Campina Grande - PB, Brazil
}

\begin{abstract}
The cosmic distance duality relation is a milestone of cosmology involving the luminosity and angular diameter distances.
Any departure of the relation points to new physics or systematic errors in the observations, therefore tests of the relation are extremely important to build a consistent cosmological framework. Here, two new tests are proposed based on galaxy clusters observations (angular diameter distance and gas mass fraction) and $H(z)$ measurements. By applying Gaussian Processes, a non-parametric method, we are able to derive constraints on departures of the relation where no evidence of deviation is found in both methods, reinforcing the cosmological and astrophysical hypotheses adopted so far.

\end{abstract}


\maketitle

\section{Introduction}

One of the key relations in cosmology is the cosmic distance duality relation (CDDR), which expresses a connection between two observables: the luminosity
distance $d_L$ and the angular diameter distance $d_A$ through

\begin{equation}
\frac{d_L}{(1+z)^2 d_A} =\eta=1 ,
\end{equation}
where $z$ is the redshift of the source.

Although it appears to be a fortuitous relation in an FLRW (Friedmann -- Lema\^{i}tre -- Robertson -- Walker) universe in many textbooks, it was actually derived by Etherington long ago \citep{etherington}, where its validity implies photon number is conserved, gravity is described
by a metric theory and light propagates on unique null geodescis \citep{bk2004}. Therefore, any violation of this relation is a clear indication
of new physics. 

Approaches to test the CDDR in the recent literature assume a cosmological model suggested by a set of observations (usually the $\Lambda$CDM concordance model) 
and check the validity of the CDDR in the context of some astrophysical effect. Examples of these methods are given by
\cite{bk2004,uzan2004,Avgoustidies}. Other tests have been proposed with the goal of testing effectively such relation, involving from luminosity distances of type Ia supernovae, angular diameter distances of galaxy clusters, gas mass fractions of galaxy clusters,  the blackness of the cosmic microwave background and gamma-ray bursts \cite[e.g.][]{uzan2004,debernardis2006,holanda2010,holanda2011,holanda2012,gha2012,ellis2013,hb2014}.  Generally, departures from the relation are written as $\eta(z) = (1+z)^{-2}d_L/d_A$, where for $\eta$ is adopted simple parametric forms as $\eta(z) = 1 + \eta_0 z$ or $\eta(z) = 1 + \eta_0 \frac{z}{1+z}$. 

An interesting test for the distance duality relation was recently proposed by \cite{hga2012}, where one can write $\eta = f_{SZE}/f_{X-ray}$,
$f_{SZE}$ standing for the gas mass fraction of a galaxy cluster measured through the Sunyaev-Zel'dovich effect \citep{sz} and $f_{X-ray}$
for the gas mass fraction obtained calculating the gas mass observed in X-rays divided by the total mass assuming hydrostatic equilibrium.
The method has some advantages comparing to other approaches: as the same object is used to determine $\eta$, there are no problems of comparing
distances in completely different regions of the sky where inhomogeneities can be important, as well as it is easier to grasp systematic errors.
Considering the two standard parametrizations for $\eta$ described above, no violation for the relation was the derived for a subset 
well described by hydrostatic equilibrium. 

In this paper, two new tests are proposed based on the distance to galaxy clusters from the ESZ/X-ray technique, X-ray gas mass fraction and $H(z)$ measurements. 
The use of $H(z)$ measurements allows a model-independent test, overcoming difficulties when using galaxy cluster and type Ia supernovae (SNe Ia) due to a dependence on a cosmological model and the Hubble constant when using SNe Ia \citep{Yang2013}.
Also, another advantage of our tests is that they rely on data coming from the observation of the same galaxy clusters, which allow us to search for new physics as well as to check the consistency of the hypotheses adopted in such observations. As we shall see, the results are derived using Gaussian Processes, a non-parametric
method which does not restrict a functional form for $\eta(z)$, where both methods fully agree with the standard value $\eta=1$ in all redshift range covered by data.
The paper is organized as follows: in Sec. \ref{samples} we describe the sample used in the work. Sections \ref{methods} and \ref{gp} details the methods and the non-parametric
approach adopted, respectively. The results are in Sec. \ref{results}. We close the paper in Sec. \ref{conclusions} with the conclusions.

\section{Samples}
\label{samples}

In order derive constraints for $\eta$ free from parametrizations we use:

(a) 38 gas mass fraction of galaxy clusters from \cite{LaRoque} spanning redshifts from $0.14$ to $0.89$, derived from Chandra X-ray data. These authors used the non-isothermal double $\beta$-model for gas distribution. The 3D temperature profile is modeled assuming that the ICM is in hydrostatic equilibrium with a NFW dark matter density distribution \citep{Navarro1997}.   

(b) 38 angular diameter distance  from galaxy clusters obtained from their Sunyaev-Zeldovich and X-ray observations (the so-called SZE/X-ray technique). The sample is in redshift range is $0.14<z<0.89$ and was compiled by \cite{Bonamente2006}  where the cluster plasma and dark matter distributions were analyzed assuming  a non-isothermal spherical double $\beta$ model. The galaxy clusters of this sample are exactly the same of \cite{LaRoque}.

(c) 28 Hubble parameter versus redshift data points, $H(z)$,  from cosmic chronometers, BAOs and a local measurement in redshifts up to  $ z = 1.75$ \citep[][]{Simon2005,Gaztanaga2009,Stern2010,Riess2011,Moresco2012,
Blake2012,Zhang2014}. For a compilation of $H(z)$ measurements see e.g. \citep{Farooq2013}.

\section{Methods} 
\label{methods}

While the item (c) in the earlier section is insensitive with respect the cosmic duality relation validity, the astronomical observations in items (a) e (b)  are dependent  through different ways. In our analysis we use two methods to obtain constraints on $\eta$, namely:

\subsection{Method I}

The gas mass fraction of galaxy clusters is given by \citep{Sasaki1996}
\begin{equation}
\label{final}
f_{gas} \equiv \frac{M_{gas}}{M_{tot}} \propto d_L{d_A}^{1/2}\;,
\end{equation}
where $M_{gas}$ is the gas mass and $M_{tot}$ is the total mass (including dark matter) of the galaxy cluster. As it is known, along with the assumption of a constant $f_{gas}$ with redshift  it is possible  to test different cosmological scenarios. However, in a recent paper, \cite{gha2012} showed that the gas mass fraction is dependent on $\eta$ value. These authors deduced a more general expression for the gas mass fraction  given by
\begin{equation}
f_{gas}(z) = N \left[\frac{d_L^{*}{d_{A}^{*1/2}}}{d_L{d_{A}^{1/2}}}\right]\;.
\end{equation}
 By using a general version of equation (1) ($\eta\neq 1$) was obtained:
\begin{equation}
\label{GasFrac4}
d_A(z) = N^{2/3} \left[\frac{{d_{A}^{*}}}{\eta^{2/3}{f_{gas}}^{2/3}}\right],
\end{equation}
where $N$ is the normalization factor that carries all the information about the matter content in the cluster and the quantities with the subscript * correspond to those used in the observations (concordance model where $\eta=1$). In this way, it is possible to define
\begin{equation}
\label{etateste}
\eta_{obs}(z)=N\left[\frac{d^{*3/2}_A}{d^{3/2}_Af_{gas}(z)}\right]\;.
\end{equation} 
In our analysis, $d_A$ is obtained directly from $H(z)$ measurements, as it will be explained later. It is important to stress that the test for $\eta$ proposed here is independent of the assumption of a constant $f_{gas}$ with redshift since we are interested in $d_A$ from each cluster separately and not from its distribution over $z$. The sources of uncertainty in the measurement of $f_{gas}$ to the galaxy cluster samples are: i) the statistical errors concerning to the X-ray and SZE observations comprehend known effects:  SZE point sources $\pm 4\%$,  kinetic SZ $\pm 8\%$, $\pm 20\%$  and $\pm 10\%$ for cluster asphericity to $f_{gas}$ from X-ray and SZE observations, respectively, ii) the systematic errors for the galaxy clusters are:  X-ray absolute flux calibration $\pm 6\%$, X-ray temperature calibration $\pm 7.5\%$, SZE calibration $\pm 8\%$  and a one-sided systematic uncertainty of $-10\%$ to the total masses which accounts for the assumed hydrostatic equilibrium  (see subsection 5 in \cite{LaRoque}). In our analysis the systematic errors will be add in quadrature to statistical errors \citep{dagostini}.

\subsection{Method II}

As it is largely known, the X-ray emission of the intracluster galaxy medium can be combined with the Sunyaev-Zeldovich effect to estimate
the angular diameter distance (ADD). The main advantage of this method is being fully independent of any local calibrator. In the context of this phenomenon, the angular diameter distance is such that
\begin{equation}
d_{A}(z)\propto \frac{{d_A}^2(\Delta
T_{0})^{2}\Lambda_{eH0}}{{d_L}^{2}
S_{X0}{T_{e0}}^{2}}\frac{1}{\theta_{c}}\propto \frac{(\Delta
T_{0})^{2}\Lambda_{eH0}}{(1+z)^4
S_{X0}{T_{e0}}^{2}}\frac{1}{\theta_{c}},
\end{equation}
where $S_{X0}$ is the central X-ray surface brightness, $T_{e0}$ is
the central temperature of the intra-cluster medium, $\Lambda_{eH0}$
is the central X-ray cooling function of the intra-cluster medium,
$\Delta T_0$ is the central decrement temperature, and $\theta_{c}$
refers to a characteristic scale of the cluster along the line of
sight (l.o.s.), whose exact meaning depends on the assumptions
adopted to describe the galaxy cluster morphology.

In  \cite{uzan2004} it was shown that this technique is heavily dependent on the cosmic distance duality relation. If one assumes a deformed expression such as equation (1) ($\eta \neq 1$) it is possible to obtain that 
\begin{equation}
d_{A}^{\: data}(z)=d_{A}(z)\eta(z)^{2}. \label{recc}
\end{equation}
This quantity is reduced to the standard angular diameter
distance only when the CDDR relation is strictly valid ($\eta=1$). In this way, it is possible to define
\begin{equation}
\label{etateste2}
\eta_{obs}(z)=\left(\frac{d_{A}^{\: data}(z)}{d_{A}(z)}\right)^{1/2}\;.
\end{equation}
In our analysis, $d_A(z)$ is obtained directly from $H(z)$ measurements. The sources of uncertainty in the measurement of $d_{A}^{\: data}$(z) to the galaxy cluster samples are: i) statistical contributions: SZE point sources $\pm 8$\%, 
X-ray background $\pm 2$\%, Galactic N$_{H}$ $\leq \pm 1\%$, $\pm 
15$\% for cluster asphericity, $\pm 8$\% kinetic SZ and for CMB 
anisotropy $\leq \pm 2\%$, ii) estimates for systematic effects are as 
follow: SZ calibration $\pm 8$\%, X-ray flux calibration $\pm 5$\%, 
radio halos $+3$\% and X-ray temperatute calibration $\pm 7.5$\%  (see table 3 in \cite{Bonamente2006}). 

\subsection{Distances from H(z) measurements}

{ The $H(z)$ measurements were derived from the following physical observables: cosmic chronometers \cite{Simon2005,Stern2010,Moresco2012,Zhang2014},
baryon acoustic oscillations obtained from the clustering of galaxies \cite{Blake2012,Gaztanaga2009} and from the Wide Field Camera 3 (WFC3) on the Hubble Space Telescope (HST) \cite{Riess2011}.}

{ The method of cosmic chronometers \cite{Jimenez2002} is based on the relative age of a galaxy population at different redshifts. As $H(z) = -\frac{1}{1+z}\frac{dz}{dt}$, for spectroscopic redshifts with interval $\Delta z$, a measured age difference $\Delta t$ provides a direct inference of $H(z)$. Note that this measurement is independent of absolute ages as well as the metric considered. The error budget considered is a combination of statistical and systematic errors, whereas the former is due to the calibration relation to infer the ages and the latter is due to metallicity,
stellar population synthesis model and star formation history. Other sources of systematic errors are not relevant given the size of the other errors, as progenitor bias, initial mass function and $\alpha$-enhancement.}

{The values of $H(z)$ derived by \cite{Gaztanaga2009} come from the baryon acoustic oscillations (BAOs), while the ones obtained by \cite{Blake2012} come from combining the BAOs with the Alcock-Paczinsky test \cite{APtest}.
The BAOs come from the sound waves propagating in the photon-baryon fluid in the early Universe. After the decoupling, these modes were frozen
and a preferred scale was imprinted on the distribution of matter, which can be detected analyzing the distribution of objects as galaxies.
This scale is the sound horizon at the baryon drag epoch, so it can be determined from observations of the cosmic microwave background radiation.
By measuring this scale, which is a standard ruler, at transversal and radial directions, one is able to derive the angular diameter distance and the Hubble parameter, respectively. The Alcock-Paczinsky test relies on comparing the radial and transversal sizes of objects assumed to be isotropic. 
Equating their sizes, one is able to derive constraints on a combination of angular diameter distance and Hubble parameter independently of a standard ruler. Both tests are weakly dependent on the adopted cosmological model, therefore it can be used in cosmological tests. 
\cite{Gaztanaga2009} estimated the impact from systematic errors in the measurement of the correlation function as well as the BAO peak location, where simulations were extensively used. Possible sources of systematics include the choice of radial selection function, finite volume effects and the accuracy of the statistical error model. In this work we considered the statistical and systematic errors obtained in \cite{Gaztanaga2009}. \cite{Blake2012} discarded possible sources of systematic errors as much smaller than statistical errors, as modelling redshift-space distortions, varying fitting range for the BAO peak and implementation of a quasi-linear model. Also, comparison with constraints coming from simulations found no evidence of systematic errors.}

We follow the methodology firstly presented by \cite{hca2013} where one transforms  $H(z)$ measurements into cosmological distance estimates by solving numerically the comoving distance integral for non-uniformly spaced data, i.e.,
\begin{equation}
\label{eq2}
D_C =c \int_0^z{dz^\prime \over H(z^\prime)}\approx {c\over 2}\sum_{i=1}^{N} (z_{i+1}-z_i)\left[ {1\over H(z_{i+1)}}+{1\over H(z_i)} \right],
\end{equation}
Since the error on $z$ measurements is negligible, we only take into account the uncertainty on the values of $H(z)$ { (the error from the method itself is also completely negligible compared to the errors of $H(z)$)}. As one may check, by using standard error propagation techniques, the error associated to the $i^{th}$ bin is given by
\begin{equation}
s_i={c\over 2}(z_{i+1}-z_i)\left({\sigma_{H_{i+1}}^2\over H_{i+1}^4} + {\sigma_{H_{{i}}}^2\over H_{i}^4}\right)^{1/2}\;,
\end{equation}
so that the error of the integral (\ref{eq2}) in the interval $z=0$ -- $z_{n}$ is $\sigma^2_n = \sum_{i=1}^n s_i$.  Thus,  we use 28 independent $H(z)$ measurements
\citep{Simon2005,Stern2010,Moresco2012,Zhang2014,Blake2012,
Gaztanaga2009,Riess2011} in the redshift range $ 0 < z < 1.75$. 


As one may see, the comoving distance obtained from this technique is completely independent of the cosmic distance duality relation. We restrict our analysis to the flat case,  in this way the ADD in methods (I) and (II) is $d_A=(1+z)^{-1}D_C$. Moreover, since the galaxy clusters and $H(z)$ observations are performed at different $z$, we calculate $D_C$ at each galaxy cluster redshift from a  polynomial fit of the $D_C$ from $H(z)$ points shown in Fig. (1). In Fig. (2) we show the $\eta_{obs}$ values for each method by using equations (5) and (8). For method (I) we marginalize over the parameter $N$. It is important to stress that all systematical errors are being considered.

\section{Gaussian Processes}
\label{gp}

Gaussian Processes (GPs) is a non-parametric method which allows one to reconstruct a function without specifying a form for it.
GPs generalize the notion of a gaussian random variable for a random function, being characterized by a mean and a covariance function.
Our knowlegde about the properties of the function are given in the covariance function, where we adopt the standard squared exponential covariance function $k(z,\tilde{z})$:

\begin{equation}
k(z,\tilde{z}) = \sigma_f \exp{\left[-\frac{(z-\tilde{z})^2}{2l^2}\right]},
\end{equation}
where $\sigma_f$ and $l$ are hyperparameters which control how the function changes in the $y$ and $x$ axes, respectively, and the covariance function relates to points in input space $z$ and $\tilde{z}$. By considering that
the function we want to reconstruct is a given realization of a GP, we have to calculate the probability of observing
a new set of points at $\mathbf{z^*}$ with mean values $\bar{\mathbf{f^*}}$ and covariance $cov(\mathbf{f^*})$ given we measured $\mathbf{y}$
at positions $\mathbf{z}$ with covariance matrix $C$, where the mean and covariance are given by

\begin{equation}
\bar{\mathbf{f^*}} = K(\mathbf{z^*},\mathbf{z})[K(\mathbf{z},\mathbf{z}) + C]^{-1}\mathbf{y},
\label{eq_mean}
\end{equation}

\begin{equation}
cov(\mathbf{f^*}) = K(\mathbf{z^*},\mathbf{z^*}) - K(\mathbf{z^*},\mathbf{z})[K(\mathbf{z},\mathbf{z}) + C]^{-1}K(\mathbf{z},\mathbf{z^*}).
\label{eq_cov}
\end{equation}
In the above expressions $K(\mathbf{z_1},\mathbf{z_2})$ is a matrix connecting all possible values of $\mathbf{z_1}$ and $\mathbf{z_2}$.
The hyperparameters must be marginalized over, where we follow the approach described in \cite{seikel2012}.
GPs have been adopted in many cosmological applications
\citep[e.g.][]{seikel2012b,landman2014,yahya2014,julien,busti2014,nair2014,busti2015,javier2015,tao2015}.

\section{Results}
\label{results}

In order to derive constraints to $\eta(z)$ two steps are needed. Firstly, we derive $\eta$ for each cluster and its error for Methods
I and II. 
From these values we are able to reconstruct $\eta(z)$ by applying a squared
exponential covariance function and imposing the condition that $\eta(0)=1.0$. Figure \ref{Fig3} and Fig. \ref{Fig4} show the constraints on $\eta(z)$ for methods I and II, respectively.
The shaded contours represent 68\% and 95\% confidence regions and all known sources of statistical and systematic errors were taken into account in our analyses. For comparison it is also plotted the results of a linear fit $\eta(z)=1+\eta_0 z$ with the 95\% confidence levels as red dashed lines.

\begin{figure}
\begin{center}
\includegraphics[width=0.45\textwidth]{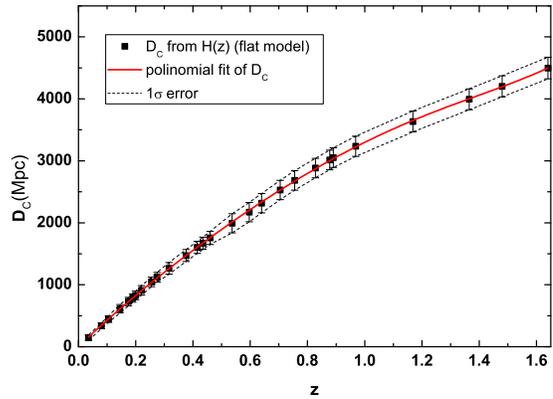}
\end{center}
\caption{The black squares points are the comoving distance obtained from H(z) measurements. The red line solid correspond to polynomial fit and the dashed lines the 1$\sigma$ error.}
\label{Fig1}
\end{figure}

\begin{figure}
\begin{center}
\includegraphics[width=0.45\textwidth]{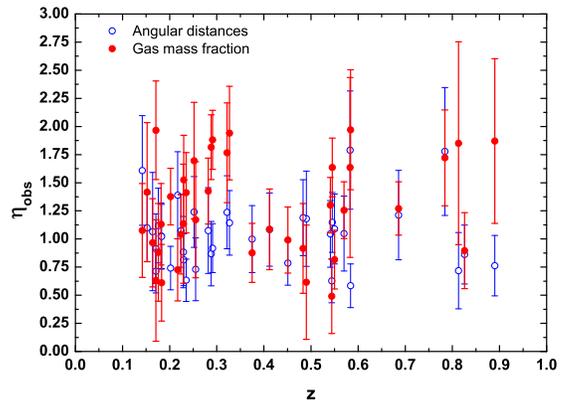}
\end{center}
\caption{$\eta(z)$ values. The filled red circles and the open blue circles are the results from Method I and II, respectively, considering all systematic errors.}
\label{Fig2}
\end{figure}

\begin{figure}
\begin{center}
\includegraphics[width=0.45\textwidth]{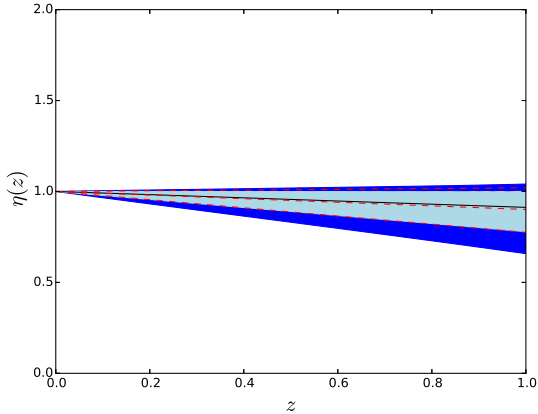}
\end{center}
\caption{Constraints to $\eta(z)$ from Method I. The black line corresponds to the mean reconstructed function and the shaded
contours 68\% and 95\% confidence levels. All known sources of systematic errors were included in the analysis and the standard
value $\eta=1$ is within the 68\% for the whole redshift region. The red dashed lines represent the mean value and 95\% confidence levels
of a linear fit $\eta(z) = 1+ \eta_0 z$.}
\label{Fig3}
\end{figure}

\begin{figure}
\begin{center}
\includegraphics[width=0.45\textwidth]{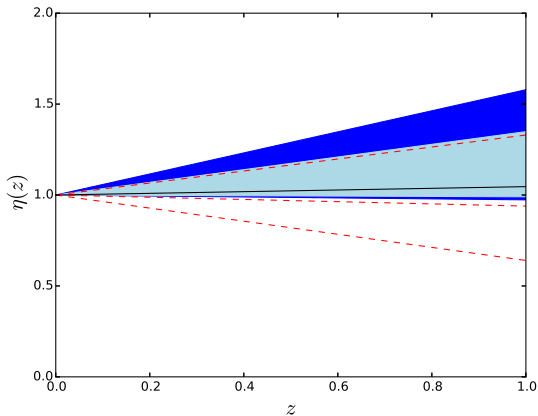}
\end{center}
\caption{Constraints to $\eta(z)$ from Method II. As in Fig. \ref{Fig3} the black line shows the mean reconstructed function and the
shaded contours represent 68\% and 95\% confidence levels, where the red dashed lines depict the best fit and 95\% confidence levels of the linear fit. 
Again all known sources of systematic errors were considered and $\eta=1$ is within the 68\% confidence level. Method II provided smaller errors than Method I, although both methods show no deviation of the cosmic distance duality relation.}
\label{Fig4}
\end{figure}

As one can see, both methods are compatible within the 68\% confidence level with $\eta=1$, i.e. with the validity of the CDDR.
Apart from not showing any evidence for new physics, our results also point to an internal consistency for deriving constraints from galaxy clusters with the gass mass fraction and with the angular diameter distance. 

In order to see whether the results were consistent given the adopted method, we redid our analysis assuming different covariance functions for the GP regression. We considered some instances of the Mat\'{e}rn family, see \citep{seikel2013} and \citep{busti2014b} for definitions and discussion.
The results were practically indistinguishable from the standard squared exponential covariance function, which reinforces the robustness of the results.

\section{Conclusions}
\label{conclusions}

In this work we have showed that observations of galaxy clusters  jointly with cosmic expansion rate can be used to derive constraints to the distance duality relation (CDDR), $\frac{d_L}{(1+z)^2 d_A} =\eta=1$. From galaxy clusters observations we used, separately, a sample of 38 angular diameter distances obtained via their gas mass fraction in X-ray band \citep{LaRoque}  and another one of 38 angular diameter distances obtained via their Sunyaev-Zeldovich effect and X-ray observations \citep{Bonamente2006}. The galaxy clusters are the same in the samples and they are in redshift range $0.14<z<0.89$.  As stressed in this paper, both observations are heavily  dependent on the CDDR validity. In order to obtain CDDR independent  distances  and perform our analyses we have used 28 recent $H(z)$ measurements. The $H(z)$ measurements are transformed into cosmological distance estimates by solving numerically the comoving distance integral for non-uniformly spaced data (see Fig. (1)). Unlike what happens in most  papers in the literature no parametrization was used to $\eta$, where we used  Gaussian Processes, a non-parametric method to reconstruct the $\eta$ function. As a general result we obtained that validity of the relation was ensured  within $1\sigma$ for the redshift range $0.14 < z <0.89$ including all systematic errors to both galaxy clusters observations. Since some Sunyaev-Zeldovich surveys are underway, or planned to start in the near future,  more and larger data sets with smaller statistical and systematic uncertainties will be available and the method proposed here may improve the constraints on fundamental physics as well as the astrophysics of galaxy clusters.

\acknowledgments

The authors would like to thank H.L. Bester for providing his GP code.
VCB is supported by CNPq-Brazil, with a fellowship within the program Science without Borders, FAPESP (grant number 2014/21098-1) and CAPES.
RFLH is supported by INCT-A and CNPq (No. 478524/2013-7).

\end{document}